\documentclass{article}
\usepackage[utf8]{inputenc}
\usepackage{graphicx}
\usepackage{url}
\usepackage[round, sort&compress, numbers]{natbib}
\usepackage{setspace}
\usepackage[margin=1in]{geometry}
\usepackage{amsmath}
\usepackage{mathabx} 
\usepackage{authblk}

\newlength{\normalwidth}
\newlength{\reducedwidth}
\newlength{\morereducedwidth}
\title{Quantifying the Vulnerabilities of the Online Public Square \\ to Adversarial Manipulation Tactics}

\author{\normalsize{Bao Tran Truong$^*$, Xiaodan Lou, Alessandro Flammini, Filippo Menczer\\
Observatory on Social Media, Indiana University, 1015 E 11th St, Bloomington, IN 47408, USA}}

\date{}

\begin{document}
\maketitle

\def\thefootnote{*}\footnotetext{To whom correspondence should be addressed: baotruon@iu.edu}

\def\thefootnote{\arabic{footnote}}

\setcounter{footnote}{0}

\newcommand{\SI}{Supplementary Material}
\newcommand{\indep}{\rotatebox[origin=c]{90}{$\models$}}

\newcommand{\SimSoM}{\emph{SimSoM}}

\begin{abstract}
Social media, seen by some as the modern public square, is vulnerable to manipulation. 
By controlling inauthentic accounts impersonating humans, malicious actors can amplify disinformation within target communities. 
The consequences of such operations are difficult to evaluate due to the challenges posed by collecting data and carrying out ethical experiments that would influence online communities. 
Here we use a social media model that simulates information diffusion in an empirical network to quantify the impacts of adversarial manipulation tactics on the quality of content. 
We find that the presence of hub accounts, a hallmark of social media, exacerbates the vulnerabilities of online communities to manipulation. 
Among the explored tactics that bad actors can employ, infiltrating a community is the most likely to make low-quality content go viral. 
Such harm can be further compounded by inauthentic agents flooding the network with low-quality, yet appealing content, but is mitigated when bad actors focus on specific targets, such as influential or vulnerable individuals.
These insights suggest countermeasures that platforms could employ to increase the resilience of social media users to manipulation.
\\
\\
\textbf{Significance Statement:} 
We show that social media users are vulnerable to adversarial manipulation tactics, through which bad actors can amplify exposure to content that threatens, for example, democratic elections and public health. 
While tactics such as flooding the network with low-quality yet appealing content are damaging, getting users to follow inauthentic accounts has the most detrimental impact. Bad actors can increase harm by maximizing coverage rather than targeting particular individuals, such as influential ones. The varying degrees of harm associated with these tactics highlight tradeoffs and specific areas on which platforms could focus as they develop deterrence strategies against manipulation.
\end{abstract}

\newpage
\section{Introduction}

The vision of social media as the modern \emph{public square} has been challenged as users have become victims of manipulation by astroturf~\cite{Truthy_icwsm2011class,metaxas2012social}, trolling~\cite{stewart2018examining}, impersonation~\cite{arif2018acting}, and misinformation~\cite{fake-news-manifesto,Shao2018anatomy,Grinberg374}. False news have been reported to spread virally --- similarly to reliable information~\cite{Shao18hoaxybots} or even more~\cite{Vosoughi1146} depending on operational definitions. These kinds of manipulation exploit a complex interplay of socio-cognitive~\cite{lin2019existential,MenczerHills2020SciAm}, ideological~\cite{Grinberg374}, and algorithmic~\cite{Nematzadeh2017popularity-bias,Nikolov2018biases} biases. 
The exploitation is enabled or greatly facilitated by inauthentic accounts that impersonate people with malicious intent. Many such accounts can be coordinated by a single entity~\cite{Pacheco2021Coordinated}. If this is done through software, such accounts are commonly referred to as \emph{social bots}~\cite{socialbots-CACM,doi:10.1002/hbe2.115}.
Inauthentic and/or coordinated accounts have been observed to amplify disinformation~\cite{Shao18hoaxybots}, 
influence public opinion~\cite{FM7090,10.1371/journal.pone.0214210,ferrara2017disinformation,Pacheco2021Coordinated}, commit financial fraud~\cite{mirtaheri2019crypto,Pacheco2021Coordinated}, infiltrate vulnerable communities~\cite{stella2018bots,stewart2018examining,Caldarelli2019The-role-of-bot}, and disrupt communication~\cite{abokhodair2015dissecting,Suarez-Serrato2016}. 

Can social media be manipulated to the point that they no longer function as a public square? Under what conditions? It is difficult to carry out empirical experiments and analyses in the real world to explore these questions~\cite{centola2018experimental, monsted2017evidence}. 
One challenge is the limited size of experiments in the wild, stemming from both costs and ethical concerns about the potentially harmful nature of content from bad actors. 
A second difficulty is the limited data from social media platforms available to researchers~\cite{misinfo_data}, exacerbated by recent events such as changes at Twitter/X. 
These difficulties have led, for example, to conflicting accounts about whether disinformation campaigns on social media can sway elections~\cite{10.1257/jep.31.2.211,jamieson2018cyberwar,Badawy2018,Guesseaau4586,10.1371/journal.pone.0213500,FM10107}. 
Evidence suggests that these operations mainly impact specific vulnerable communities~\cite{Bail2020IRA,CSMaP2023Russian}. 
However, we lack a comprehensive quantitative understanding of how coordinated inauthentic tactics can disrupt online communities. This prevents the informed design of moderation or regulatory policies to protect the online public square from manipulation.

Here we introduce \SimSoM{}, a minimalistic model of a generic social media platform. The model allows us to explore scenarios in which an information-sharing network is manipulated by malicious actors controlling inauthentic accounts, and to measure the consequences of such information operations.
We assume that bad actors aim to spread low-quality information. While there are different kinds of low-quality content in reality --- disinformation, conspiracy theories, malware, or other harmful messages --- our model uses an abstract definition of low-quality content that encompasses these different types. The impact of the manipulation is measured in terms of the quality of information to which users are exposed in the network.

We find that the presence of manipulation is sufficient to suppress quality information, driving low-quality content to spread virally in the network. 
We also examine network vulnerabilities that may amplify the effects of manipulation, and evaluate the overall information quality as a result of different malicious tactics. We focus on four well-documented tactics commonly employed in influence operations across various platforms~\cite{diresta2018tactics}: (i)~infiltrating a community, for example through social bots~\cite{Shao2018anatomy,Yang2023Anatomy-AI-botnet}, follow trains~\cite{torres2022trains}, or by impersonating news outlets~\cite{diresta2018tactics,Yang2023Anatomy-AI-botnet}; (ii)~generating deceptively appealing content, such as novel narratives~\cite{Vosoughi1146} or emotional messages~\cite{ferrara2015measuring}; (iii)~flooding the network with high volumes of content by posting at high frequency to artificially inflate popularity/engagement indicators~\cite{Shao18hoaxybots,Yang2023Anatomy-AI-botnet,Grinberg2024Sci}, and possibly deleting content to avoid detection~\cite{Torres2022deletions}; and (iv)~targeting specific users, such as influential~\cite{Shao18hoaxybots} or vulnerable individuals~\cite{Shao2018anatomy}. 
Insights from analyses of these tactics are instrumental in developing countermeasures to increase the resilience of social media and their users against manipulation. We discuss mitigation steps that platforms could take and the issues that arise from regulations aimed at protecting human speech from suppression.

\section*{Results}

\begin{figure}[t] 
\centering
\centerline{\includegraphics[width=\normalwidth]{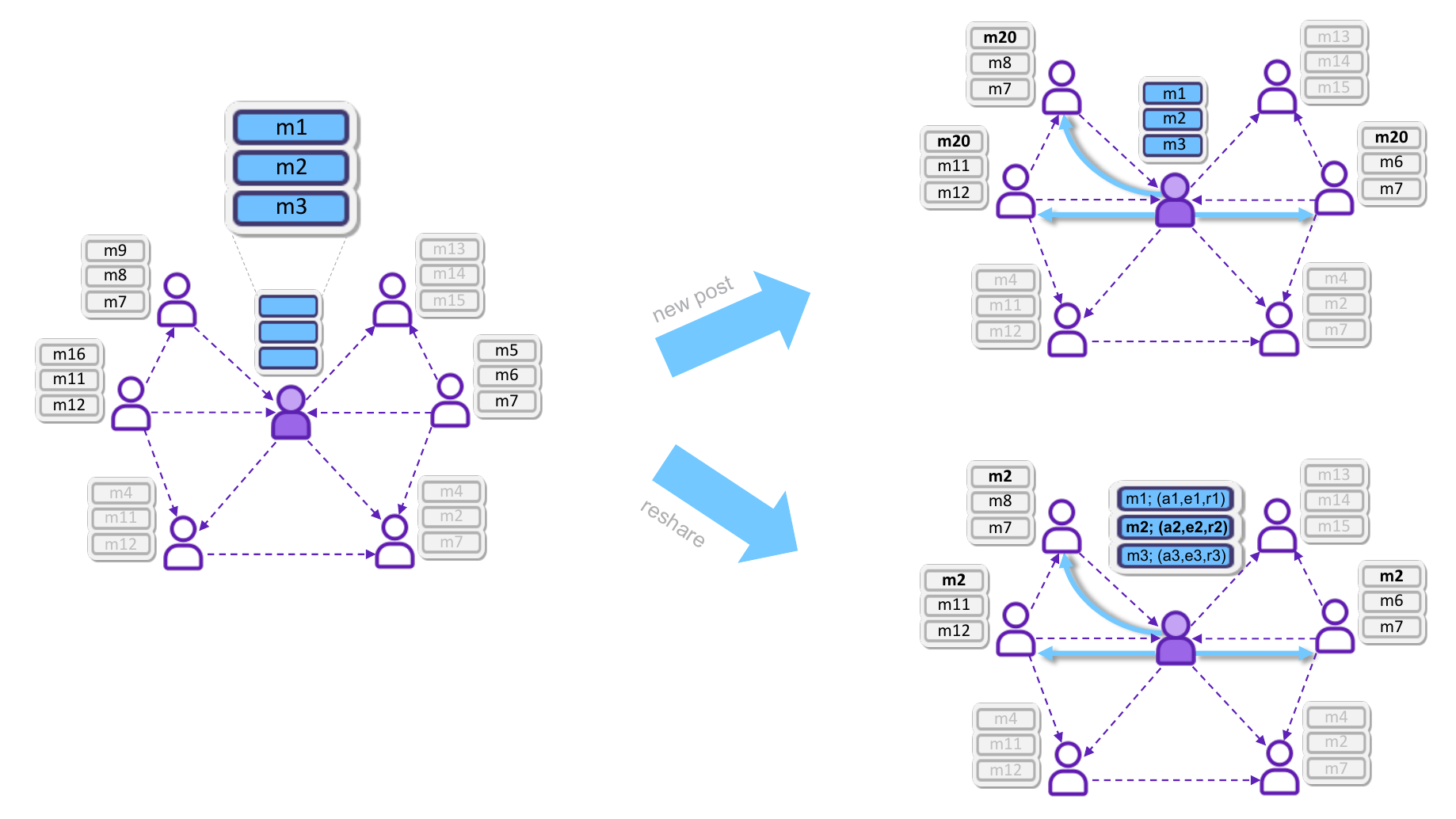}}
\caption{Illustration of the \SimSoM{} model. Each agent has a limited-size news feed, containing messages posted or reposted by friends. Dashed arrows represent follower links; messages propagate from agents to their followers along solid links. At each time step, an active agent (colored node) either posts a new message (here, \texttt{m20}) or reposts one of the existing messages in their feed, selected with probability proportional to their appeal $a$, social engagement $e$, and recency $r$ (here, \texttt{m2} is selected). The message spreads to the node's followers and shows up on their feeds.}
\label{fig:model}
\end{figure}

We model information diffusion in a social media platform such as Twitter/X, Instagram, or Mastodon. The information system is a directed network with nodes representing accounts and links representing follower relations. Similar to real-world platforms, content circulates through messages that appear in news feeds. Agents can post new messages or reshare messages from their feeds, generated by their friends, i.e., the accounts they follow.
The information diffusion process is illustrated in Fig.~\ref{fig:model}. 
Messages represent information that could take the form of text, links, hashtags, images, or other media. An agent can introduce a new message into the system 
or, alternatively, select a message from their news feed to reshare. 
Messages created and reshared by an agent then appear on the news feeds of their followers.

Even though people prefer quality content~\cite{pennycook2021shifting}, their sharing behavior is mediated by other factors such as laziness~\cite{pennycook2019lazy} and message appeal. To account for this, each message $m$ in the model has two intrinsic and independent attributes, \emph{appeal} $a_m$ and \emph{quality} $q_m$. Everything else being equal, messages with higher appeal are more likely to be reshared (Fig.~\ref{fig:model}). Quality, on the other hand, represents objective, desirable properties of content such as the originality of an idea or the accuracy of a claim.  Here we naively represent quality and appeal as scalar values. 
Deceptive posts may have low quality yet high appeal. For example, false news and junk science articles have low quality --- most people would not share them knowingly. Yet such low-quality content may be even more likely to spread virally than high-quality information~\cite{Vosoughi1146}. 
Low-quality content may be novel, clickbait, ripped from headlines, and/or may appeal to people's political, emotional, or conspiratorial bias. Worse yet, bad actors can employ generative AI to produce such content at scale~\cite{Yang2023Anatomy-AI-botnet}. 

The model captures bias towards appeal as well as two other ingredients that are prioritized by the ranking algorithms of social media platforms, namely social engagement and recency~\cite{twitteralgo}. 
An agent selects a message to reshare from the news feed, which is an inventory of distinct messages recently shared by the agent's friends. The message is selected with probability proportional to: (i)~its appeal; (ii)~its social engagement, defined as the number of times it has been shared by the agent's friends; and (iii)~its recency, which decreases with time in the feed (see details in Methods). 

Unlike authentic agents, whose intention is to consume and share high-quality information, we define inauthentic agents as accounts that are controlled by bad (adversarial) actors to spread low-quality content among authentic agents. 
We refer to these accounts as ``bad actors" or ``inauthentic agents" throughout this paper. Such accounts may be controlled by humans (trolls), software (social bots), or a mixture (cyborgs). 
The model has three parameters to model manipulation tactics by bad actors: \emph{infiltration}, \emph{deception}, and \emph{flooding}, explained next. 

\begin{figure}
\centerline{\includegraphics[width=\reducedwidth]{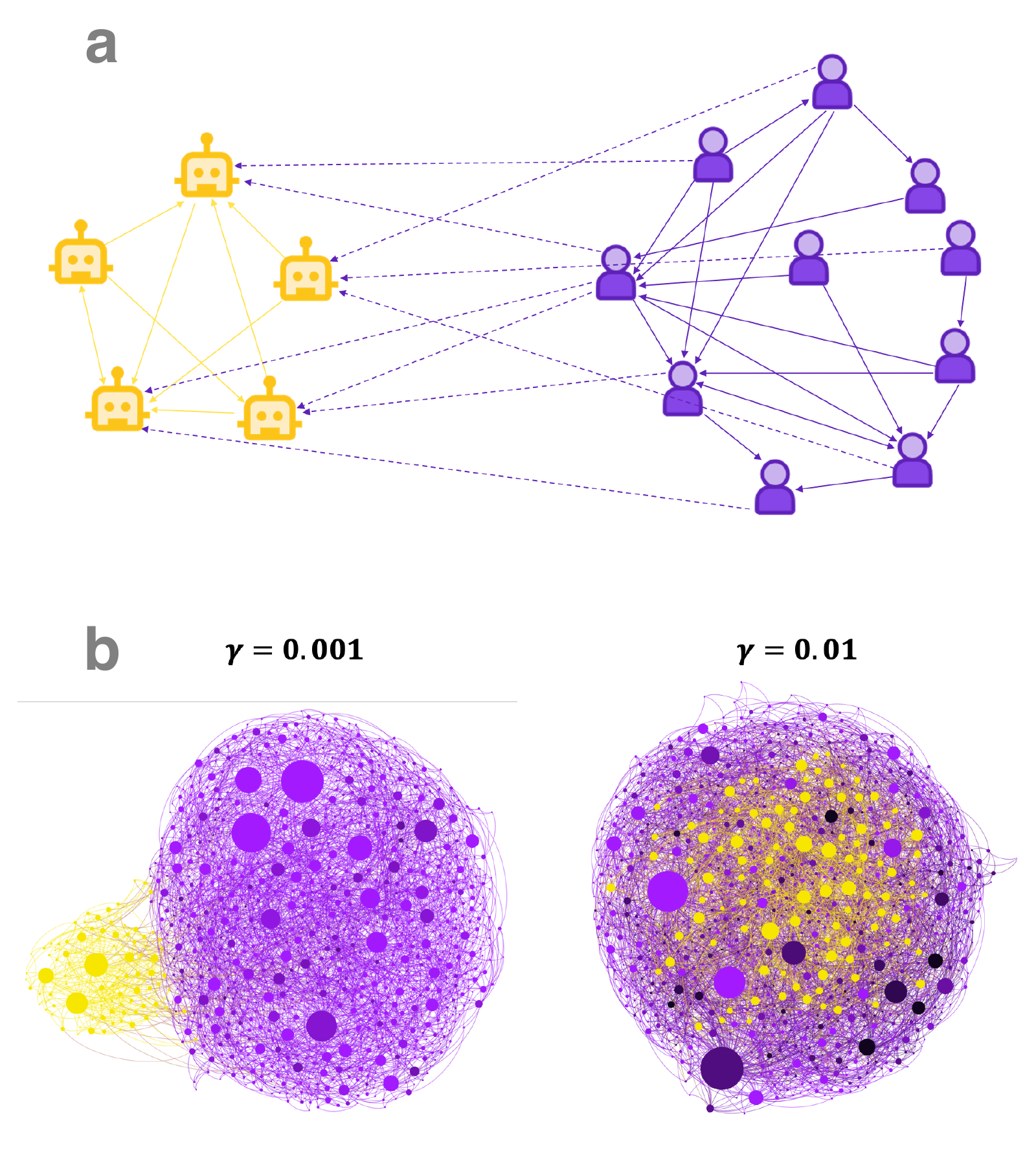}}
\caption{Subnetworks modeling authentic accounts (purple nodes) and bad actors (yellow nodes). 
(a)~Illustration of the follower link structure. Solid links indicate follower relations within each subnetwork. Both subnetworks have hub and clustering structure that mimics or derives from online social networks. Dashed links represent authentic accounts following bad actors, according to the infiltration parameter $\gamma$, which represents the probability that an authentic node follows any given bad actor. When $\gamma=0$ there is no infiltration and bad actors are isolated, therefore harmless; the opposite extreme $\gamma=1$ indicates complete infiltration, such that bad actors are followed by all authentic accounts. 
(b)~Effects of bad-actor infiltration $\gamma$ on the quality of messages in synthetic networks with $10^3$ authentic agents and 100 inauthentic agents. 
For illustration purposes, both the authentic and inauthentic subnetworks in this panel are generated with the same method used for the inauthentic subnetworks in our experiments (see Methods). 
Node size represents the number of followers. The darker an authentic agent node, the lower the quality of messages in their feed.}
\label{fig:infiltration}
\end{figure}

Infiltration describes how bad actors amplify exposure to their messages by getting authentic accounts to follow them (Fig.~\ref{fig:infiltration}(a)). Bad-actor infiltration of the social network is modeled by a parameter $\gamma$, the probability that each bad actor is followed by each authentic agent. 
Unless otherwise stated, we assume that authentic agents follow bad actors uniformly at random. Fig.~\ref{fig:infiltration}(b) illustrates the effective suppression of information quality when $\gamma$ is high. 

The quality $q$ and appeal $a$ of messages originating from authentic accounts are drawn independently from two distinct distributions, reflecting empirical evidence that these messages tend to have high quality and low appeal (see \SI). 
In contrast, we assume that bad actors can manipulate information in the network by creating messages with low quality ($q=0$) and deceptively high appeal. The appeal differential of content from bad actors is modeled by the deception parameter $\phi$. In the absence of deception ($\phi=0$), the appeal of bad-actor messages is drawn from the same distribution as those from authentic accounts. If $\phi>0$, it represents the probability that bad-actor content has $a=1$, making it irresistible (see details in Methods).

Flooding is another tactic inauthentic accounts can use to amplify their influence, by crowding out high-quality information. 
To model this, the parameter $\theta$ is defined as a boost of engagement for bad-actor content (see Methods).

\SimSoM{} lets us explore information diffusion on social media, including the properties of authentic accounts that might render them vulnerable to adversarial attacks and the effects of different manipulation tactics. In the next sections, we present the results from simulations of the model on online communities derived from an empirical follower network ($N \approx 10^4$ Twitter accounts). 
The network has both scale-free structure (hubs) and high clustering (triangles), structural characteristics that are ubiquitous in socio-technical networks. It also has a realistic community structure, with two well-separated clusters of accounts capturing political polarization (see Methods). 

Once the system reaches a \emph{steady state}, in which the message quality across the network has stabilized (see Methods), we record the mean quality of the messages in the feeds of authentic agents. These measurements are further averaged across simulation runs with the same parameters but different random seeds. 
We simulate the information diffusion process in networks with different structures to explore whether social network features render communities vulnerable to bad-actor tactics. Similarly, we evaluate the impact of these tactics by evaluating the model with varying levels of bad-actor infiltration ($\gamma$), deception ($\phi$), flooding ($\theta$), and targeting specific types of accounts.
We report on the \emph{relative quality}, defined as the ratio of the average quality of authentic agents to that of a baseline without bad actors (see Methods). 

\subsection*{Network Vulnerabilities}

\begin{figure}
\centerline{\includegraphics[width=\normalwidth]{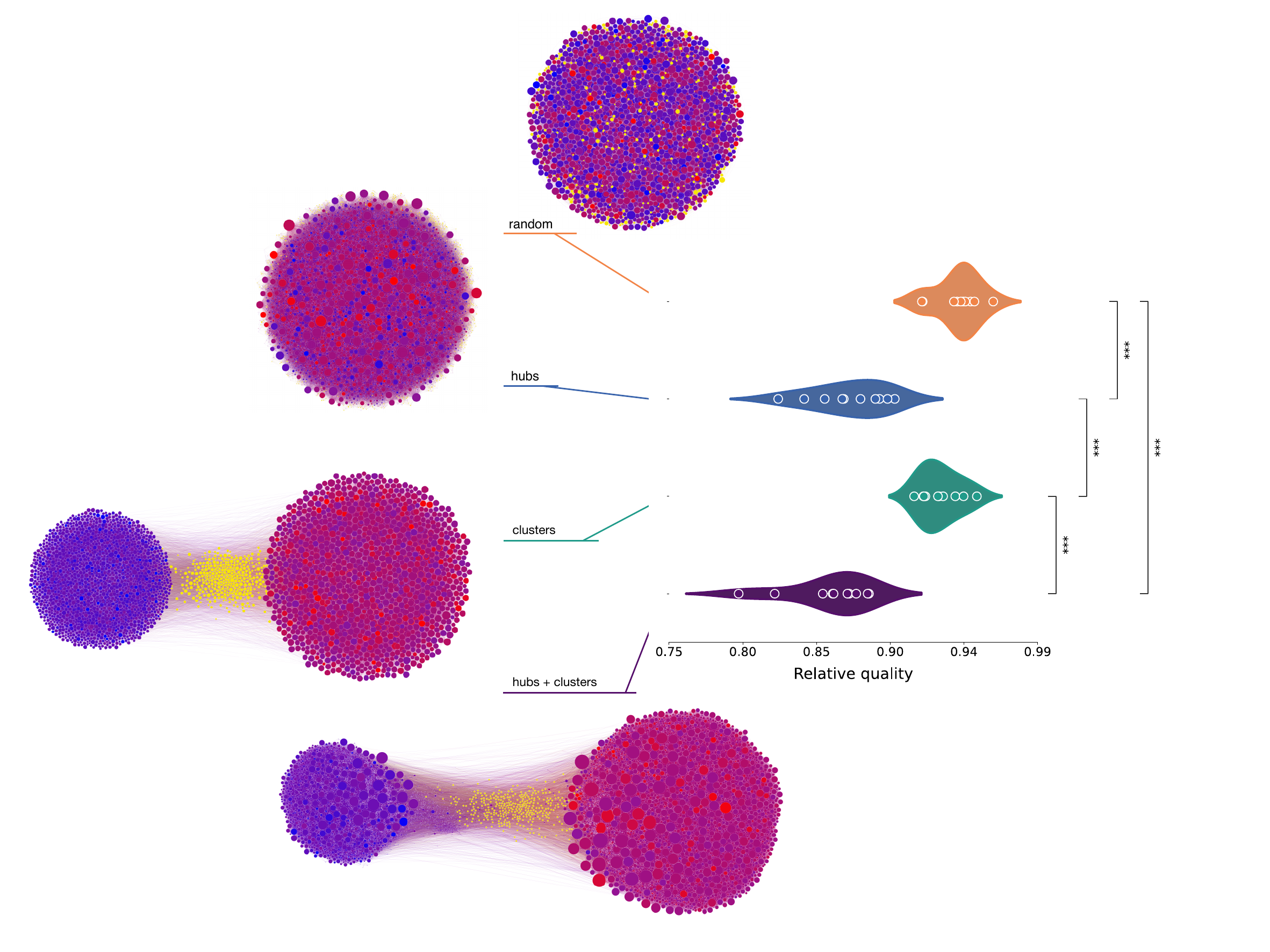}}
\caption{Impacts of different network structural features on the average information quality, relative to the scenario without bad actors. The original network (``hubs + clusters'') is visualized along with shuffled networks in which links from the original network are rewired while preserving clusters, hubs, or neither (``random''). Node size and color represent, respectively, the number of followers of an account and their political leaning ranging from liberal to conservative (red to blue, see Methods). Yellow nodes are bad actors. Pairwise statistical significance is calculated using the Mann–Whitney U test (*** for $p < 10^{-3}$); only significant differences are reported.}
\label{fig:shuffle}
\end{figure}

Key structural features of the social network may play a role in amplifying our vulnerability to manipulation by bad actors. The empirical network has two features that are ubiquitous in social media: the presence of hubs and highly clustered communities. We can explore how overall quality is affected by these features through three alternative networks constructed by shuffling links while preserving hubs, cluster structure, or neither (see Methods). 
Fig.~\ref{fig:shuffle} shows that the presence of clusters does not significantly affect the overall quality (purple vs. blue) because we assume that agents in each cluster are equally likely to follow bad actors. On the other hand, having hubs makes a network more vulnerable to manipulation: the relative system quality is significantly lower in the presence of hubs, both when there are clusters (purple vs. green, $p < 10^{-3}$) and not (blue vs. orange, $p < 10^{-3}$). This is because node in-degrees and out-degrees are highly correlated (Spearman correlation 0.9, $p < 10^{-4}$) in the empirical network. Therefore, when authentic followers are concentrated among hubs, high-quality content is also concentrated among those hubs. This implies that high-quality content has more competition and becomes obsolete more quickly compared to the case in which this content is uniformly distributed among authentic nodes. The same does not apply to content from bad actors because this content spreads uniformly among authentic users. (The scenario in which content from bad actors mostly concentrates among hubs is explored later.) 

\subsection*{Infiltration, Deception, and Flooding Tactics}

\begin{figure}
\centerline{\includegraphics[width=\normalwidth]{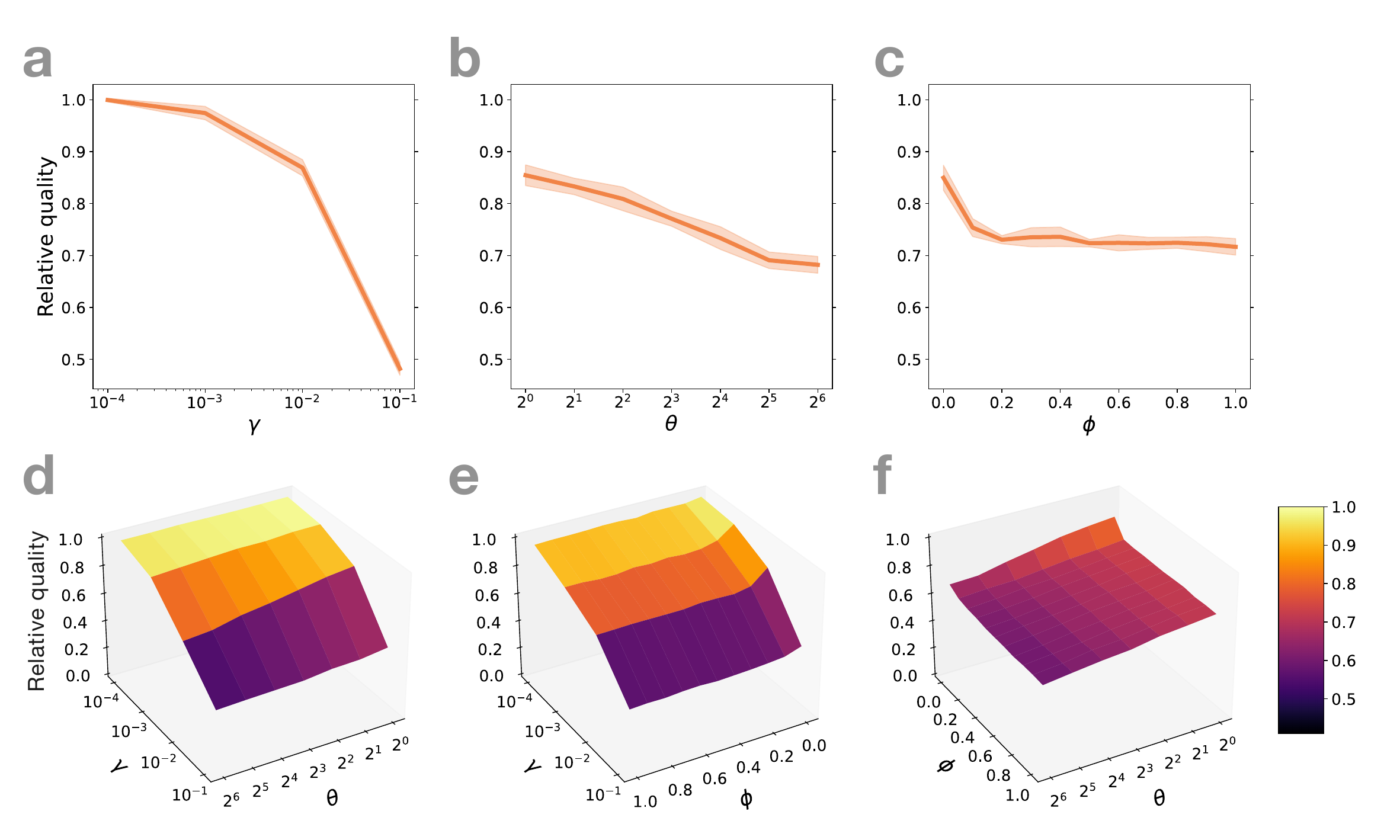}}
\caption{Effects of individual and combined tactics by bad actors on the system's message quality, relative to the scenario without bad actors. 
(a)~Varying infiltration $\gamma$, without flooding ($\theta=1$) or deception ($\phi$=0). Shading represents 95\% confidence intervals across runs in panels a--c.
(b)~Varying flooding $\theta$ with infiltration $\gamma=0.01$ and no deception ($\phi=0$).
(c)~Varying deception $\phi$ with infiltration $\gamma=0.01$ and no flooding ($\theta=1$). 
(d)~Joint infiltration and flooding with no deception.
(e)~Joint infiltration and deception with no flooding.
(f)~Joint deception and flooding with infiltration $\gamma=0.01$.
}
\label{fig:bottactics}
\end{figure}

Bad actors may maximize their message spread by combining various manipulation tactics. We systematically quantify the effects of these tactics through simulations varying the parameters for infiltration ($10^{-4} \leq \gamma \leq 10^{-1}$), deception ($0 \leq \phi \leq 1$), and flooding ($1 \leq \theta \leq 64$). See Methods for further details. Fig.~\ref{fig:bottactics}(a,b,c) illustrates the effects of individual malicious tactics on the overall quality of information spreading through the network, compared to a baseline scenario without bad actors. We observe that infiltration is the most harmful manipulation tactic: when authentic agents have a $\gamma=10\%$ probability of following each bad actor, the average quality in the system is reduced to less than half. Flooding and deception have smaller effects. When low-quality content has $\theta=64$ times more exposure than authentic content, quality is reduced to less than 70\%. When bad actors generate content with maximum appeal exclusively ($\phi=1$), quality is reduced to about 70\%. 

Similarly, Fig.~\ref{fig:bottactics}(d,e,f) shows the effects of pairs of tactics combined. Infiltration is dominant, but more harm can be done in combination with flooding or deception: the average quality is reduced to 40\% when $\gamma=0.1$ and $\theta=64$ (Fig.~\ref{fig:bottactics}(d)) or $\phi=1$ (Fig.~\ref{fig:bottactics}(e)). Combining flooding and deception ($\theta=64, \phi=1$, Fig.~\ref{fig:bottactics}(f)) only results in marginal loss of quality (below 70\%). With all three tactics combined ($\gamma=0.1, \theta=64, \phi=1$, not shown), bad actors can further reduce the quality to 30\%.

\subsection*{Reshare and Exposure Cascades}

\begin{figure}
\centerline{\includegraphics[width=\textwidth]{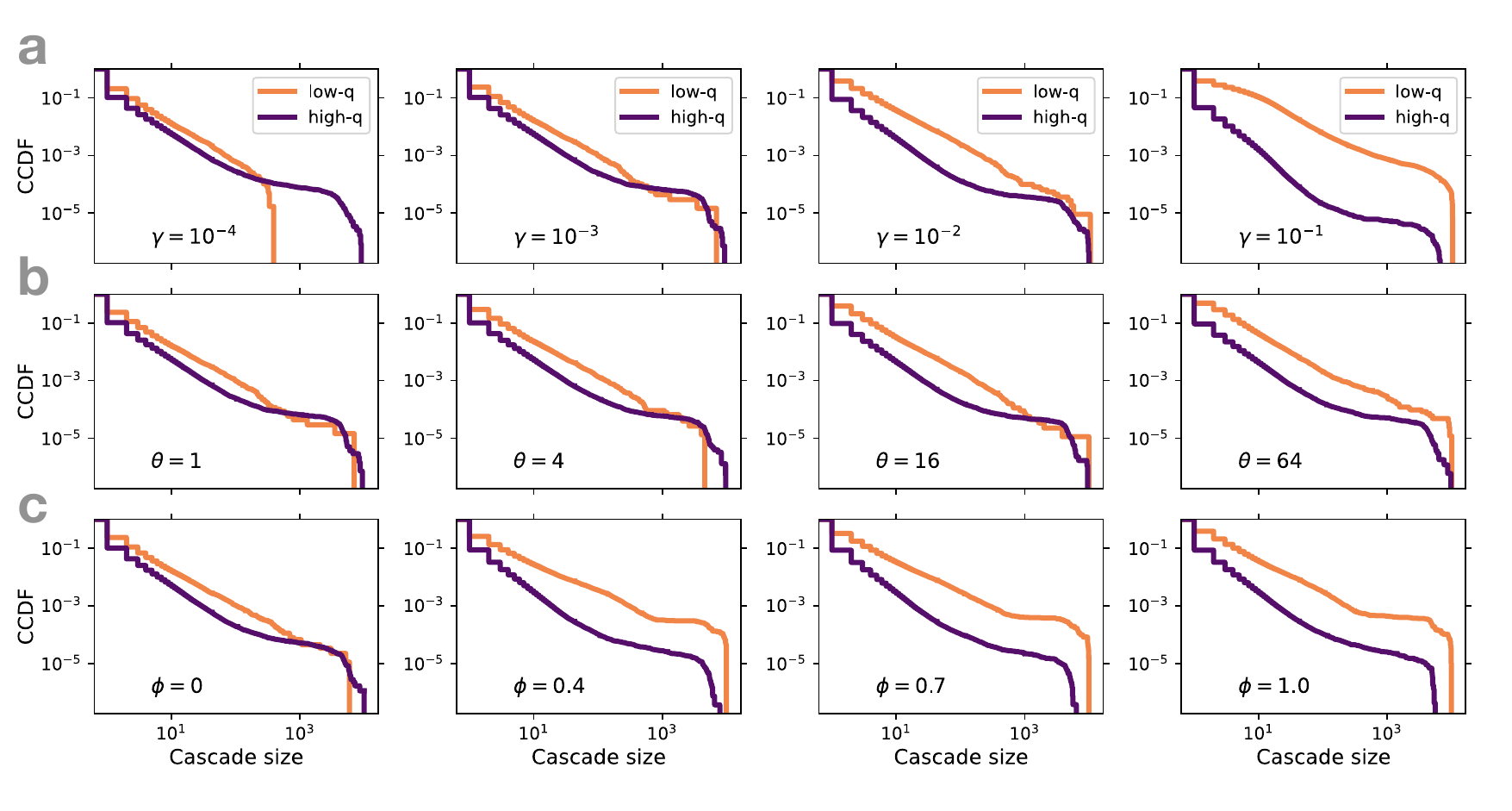}}
\caption{Complementary cumulative distributions of reshare cascade sizes for low- and high-quality content, generated by inauthentic and authentic agents, respectively. The plots are based on 10 simulations. 
(a)~Effect of bad-actor infiltration $\gamma$, with no flooding ($\theta=1$) or deception ($\phi=0$). 
(b)~Effect of flooding $\theta$, with low infiltration ($\gamma=10^{-3}$) and no deception ($\phi=0$).
(c)~Effect of deception $\phi$, with low infiltration ($\gamma=10^{-3}$) and no flooding ($\theta=1$).
}
\label{fig:cascade-distributions}
\end{figure}

Empirical evidence has shown that among fact-checked claims, low-quality content (debunked claims) tends to have larger retweet cascades than high-quality content (confirmed claims)~\cite{Vosoughi1146,Juul2021}. \SimSoM{} allows us to examine how different factors may contribute to such a virality pattern. Fig.~\ref{fig:cascade-distributions} illustrates the effects of bad-actor tactics on the size of reshare cascades for content generated by both authentic agents (``high-quality'') and inauthentic agents (``low-quality''). 
We observe that low-quality cascade size is boosted by appeal. 
This is in line with the hypothesis that factors like novelty make false news more viral~\cite{Vosoughi1146}. 
However, low-quality cascades can be boosted most effectively through high bad-actor infiltration. 
Importantly, empirical analyses that do not reconstruct actual cascades from the data may underestimate intermediary amplification by inauthentic accounts, even when those accounts are removed as originators (cascade tree roots). 
On the other hand, our model lets us reconstruct the likely reshare cascades that include intermediary amplification by inauthentic accounts (see Methods). 

\begin{figure}
\centerline{\includegraphics[width=\textwidth]{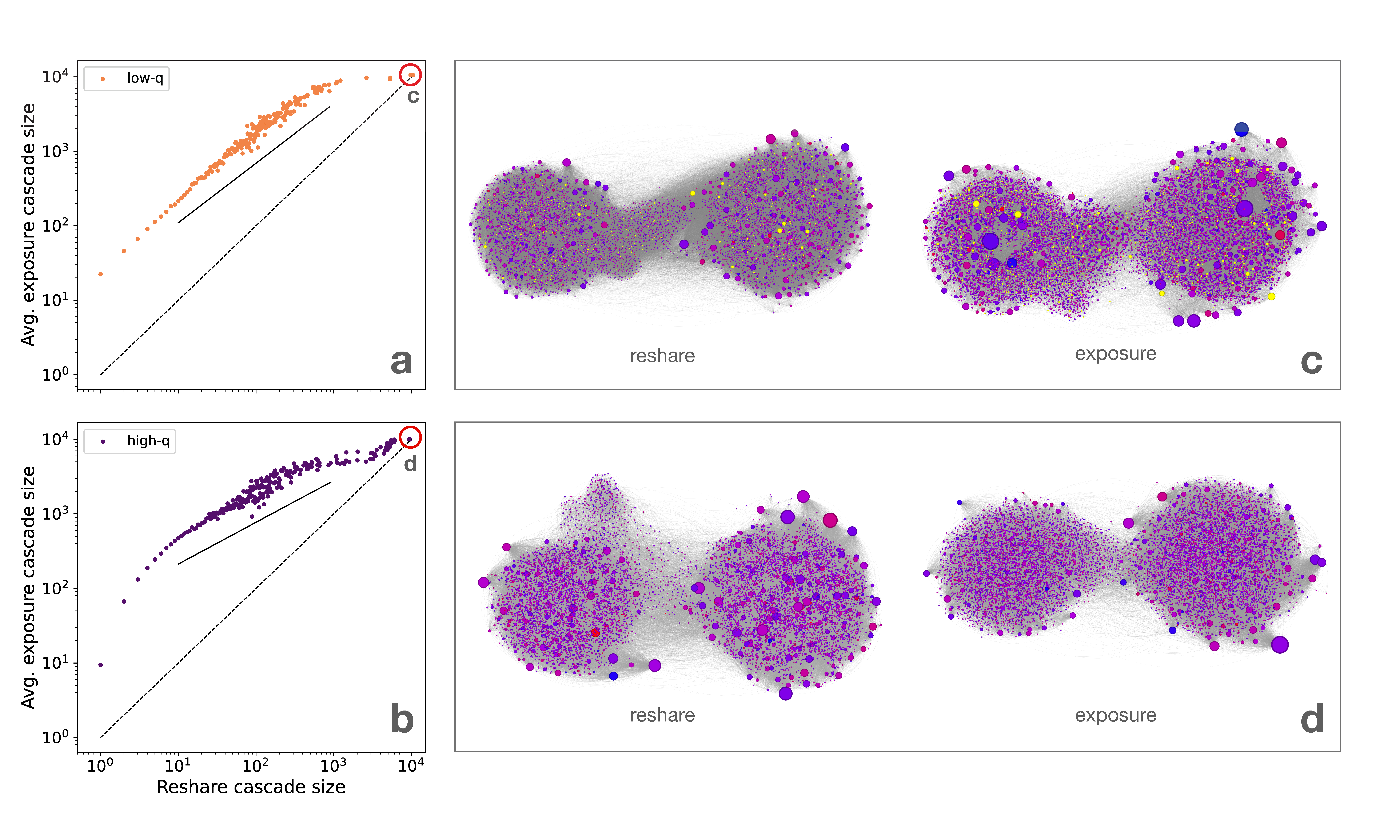}}
\caption{Scaling between reshare and exposure cascade sizes. (a)~Scaling for low-quality messages (posted by inauthentic agents). (b)~Scaling for high-quality messages (posted by authentic agents). The exposure cascade size is averaged across messages with the same reshare cascade size, based on 10 simulations. The dashed lines provide a linear scaling reference, while the solid lines show the slopes (exponents) $\nu$ of power-law fits for reshare cascades of size between 10 and 1,000, yielding $\nu = 0.80 \pm 0.01$ (low-quality messages) and $\nu = 0.56 \pm 0.01$ (high-quality messages). The largest reshare and exposure cascades (corresponding to the circles in panels a and b) are also visualized for (c)~low-quality and (d)~high-quality messages, based on one simulation. Node colors are the same as in Fig.~\ref{fig:shuffle}; node size represents out-degree, or influence. 
Here we use $\theta=1, \phi=0, \gamma=10^{-2}$; the results are similar for other $\gamma$ values. 
}
\label{fig:cascade-scaling}
\end{figure}

To date, empirical data from social media platforms has allowed researchers to measure reshare cascades, but not exposure. As a result, few studies focus on exposure~\cite{Budak2024Nat}. Using the \SimSoM{} model, it is possible to reconstruct not only likely reshare networks but also exposure networks, thus estimating how many accounts are exposed to (i.e., view) a message even if they do not reshare it (see Methods). Fig.~\ref{fig:cascade-scaling} compares the sizes of reshare and exposure cascades. 
In general, reshares underestimate exposures by roughly one order of magnitude. 
Excluding the smallest cascades and the largest ones (to avoid finite-size effects), we observe that exposure networks grow sub-linearly with reshare networks: $s_v \sim s_r^{\nu}$ where $s_v$ and $s_r$ are the exposure and reshare cascade sizes, respectively, and $\nu<1$ is the scaling exponent. This means that as messages go viral, exposures grow more slowly than reshares. 
The exponent is higher for low-quality ($\nu \approx 0.8$) than for high-quality content ($\nu \approx 0.6$), suggesting that for each extra reshare, messages posted by inauthentic agents gain more views.  

\subsection*{Targeting Tactics}

The above results show that bad actors can use inauthentic accounts to infiltrate and disrupt an online public square. We have thus far assumed that all authentic agents have the same probability of following inauthentic ones, reflecting a scenario in which bad actors do not focus their efforts on specific potential followers. However, those interested in manipulating the network may want to maximize the spread of low-quality content through the community by targeting certain groups of accounts. 

As an example, an adversary might target influentials based on the assumption that having such followers can multiply their impact --- a message reshared by an influential account has a higher chance of going viral.
Targeting influentials is well within the capability of bad actors and even automated accounts; the number of followers, often used as a proxy for influence~\cite{cha2010measuring}, is public information on all social media platforms. 
A bad actor can easily interact with accounts having many followers by mentioning and/or following them~\cite{torres2022trains,Chen2020drifters}; other ploys include retweeting, quoting, and/or liking their tweets. 
There is empirical evidence of preferential targeting by bad actors that spread misinformation~\cite{Truthy_icwsm2011class,Shao18hoaxybots}. 
Targeting politically active accounts or habitual misinformation spreaders
are also conceivable tactics. 

An important question, then, is whether tactics targeting specific authentic accounts do in fact increase the manipulative power of bad actors.
To explore this question, the model can be extended to account for various features of social media users. 
Here we consider five features that are available in the empirical data: number of followers (\emph{hubs} tactic), propensity to share misinformation (\emph{misinformation} tactic), political partisanship  (\emph{partisanship} tactic), or specific political leaning (\emph{liberal} and \emph{conservative} tactic). 
We then introduce a preferential targeting tactic for adversarial actors, according to which authentic agents have different probabilities of following inauthentic ones, proportional to one of these features. See Methods for details. 

\begin{figure}[t]
\centerline{\includegraphics[width=\normalwidth]{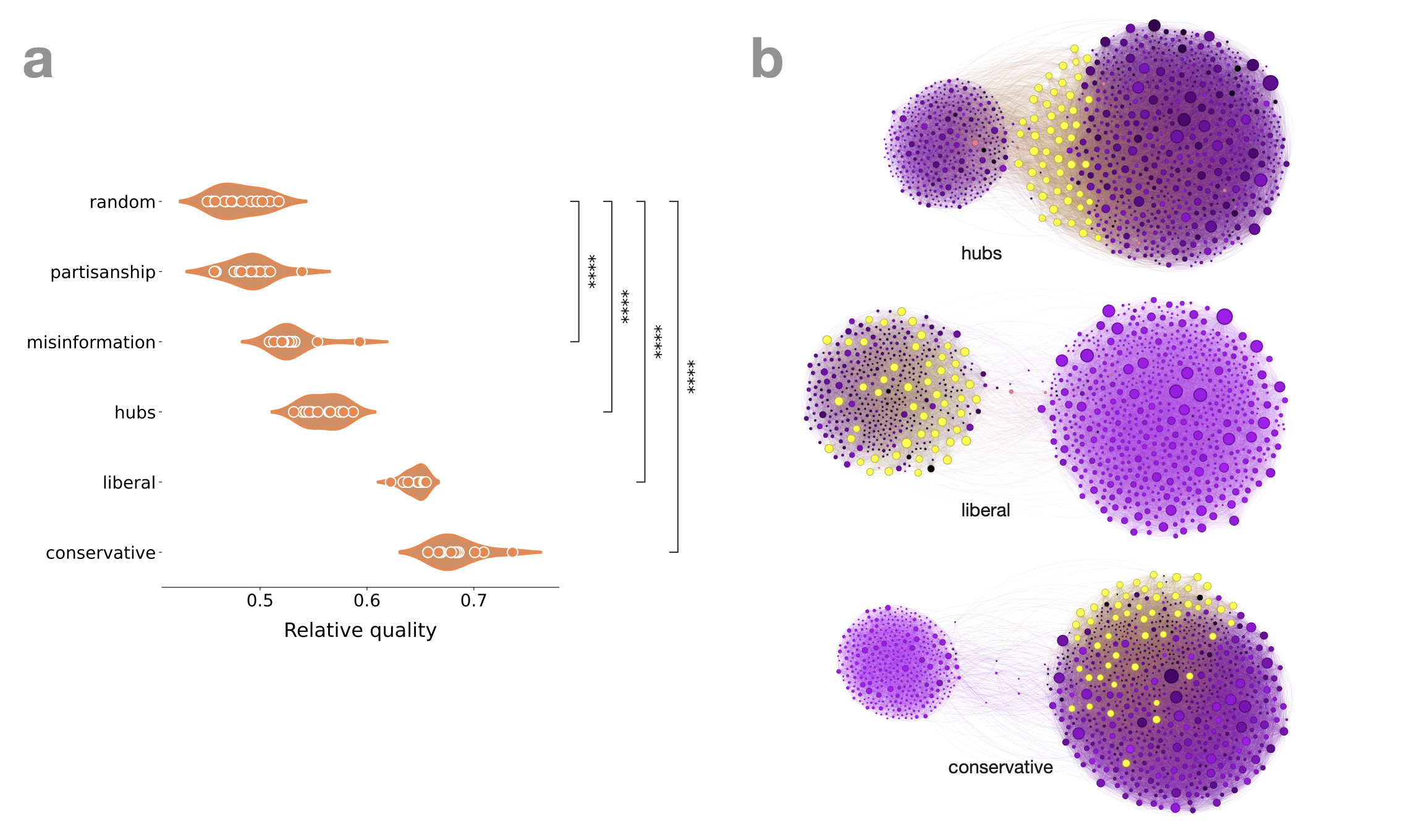}}
\caption{Effects of targeting tactics. (a)~Average information quality resulting from each tactic, as well as the default random targeting, relative to the scenario without bad actors. We highlight significant differences calculated using The Mann–Whitney U test (**** for $p < 10^{-4}$). (b)~Suppression of quality in the empirical network when bad actors specifically target influential accounts (hubs), and when they target politically left- (liberal) and right-leaning (conservative) accounts. The network has $10^3$ authentic agents (purple nodes) and 50 inauthentic agents. Node size represents the number of followers. The darker an authentic agent node, the lower the quality of messages in their feed. Significant changes due to targeting tactics are only observed when bad-actor infiltration is sufficiently high, therefore we use $\gamma=10^{-1}$ in experiments for both panels.}
\label{fig:targeting}
\end{figure}

Fig.~\ref{fig:targeting} shows the impact of these targeting tactics on information quality. 
Counterintuitively, preferential targeting is less harmful than random targeting: the distribution of quality is uneven so that the targeted population is worse off, but other parts of the community are spared. 
Targeting tactics therefore tend to backfire if we assume that bad actors intend to maximize the spread of their content across the full community.

Targeting hubs, for example, results in a network with significantly higher average quality ($p < 10^{-4}$). 
The amplification power of an influential is counterbalanced by the concentration of low-quality content, which has fewer chances to be reshared; the majority of other agents are left relatively free from manipulation. 
Note that this is analog to the reason why the presence of hubs leads to high-quality content being forgotten more quickly when hubs are not targeted by bad actors, as seen earlier. 

The harm of manipulation through the network also diminishes when bad actors target accounts sharing a lot of misinformation or with specific political leaning ($p < 10^{-4}$). 
The echo-chamber structure of the empirical network (Fig.~\ref{fig:targeting}(b)) helps interpret the latter finding: low-quality messages get shared and become obsolete rapidly within one densely connected partisan cluster, sparing the rest of the network. 

\section*{Discussion}

Social media platforms have enhanced the so-called \emph{attention economy}, in which abundant content must compete for our scarce attention~\cite{simon_designing_1971}. 
But how to ensure that accurate, relevant, timely information wins this competition? 
To date, the policies that govern social media have been mainly guided by the concept of a \emph{free marketplace of ideas}, rooted in John Milton's centuries-old reasoning that truth prevails in a free and open encounter of opinions~\cite{Milton1644Areopagitica}. 
In an ideal world, the \emph{wisdom of the crowd}~\cite{surowiecki2005wisdom} would realize this vision by combining the opinions of many users~\cite{page2008difference}. 
Unfortunately, several aspects of modern social media challenge the illusion of a public square or marketplace in which wise crowds access and select quality information~\cite{lin2019organization,RUFFO2023100531}.  
First, information production is affected by information consumption~\cite{Ciampaglia2015production}, creating incentives to produce appealing but not necessarily high-quality content.  
Second, social influence undermines the wisdom of crowds because the information and opinions to which we are exposed online may be inauthentic, correlated through coordination~\cite{Salganik854,Lorenz2011}, or dominated by few influential individuals~\cite{BeckerE5070}. 
Third, confirmation bias~\cite{nickerson1998confirmation} can increase vulnerability to misinformation in social media~\cite{HillsProliferation18,MenczerHills2020SciAm}. 
Finally, the structure of information flow networks can distort perceptions and increase vulnerability to malicious actors~\cite{gerrymandering2019,LermanParadoxDirected2020}. 

Exploration using \SimSoM{} quantifies how manipulation by bad actors can prevent social media from functioning as a public town square. 
Simulations of the model reveal that making low-quality content highly appealing plays a lesser role compared to other harmful tactics by inauthentic accounts. This suggests that novelty, for example, may not provide the primary explanation for the virality of fake news, as previously hypothesized~\cite{Vosoughi1146}.  
Tactics that inflate engagement indicators, such as flooding, can erode the system's quality just as well. 
More importantly, we find that infiltrating the network is a dominant harmful tactic available to bad actors --- these accounts can drastically suppress quality by inducing only a small fraction of the community into following them.

A wealth of previous models and experiments have focused on information diffusion and popularity. 
Several studies have investigated the role played by network mechanisms affecting the popularity of individual posts, including exogenous and endogenous bursts of attention~\cite{refcrane,Ratkiewicz10prl}, memory~\cite{refbingol}, novelty~\cite{refhuberman,wuhuberman2007}, and position bias~\cite{Hodas12socialcom,Kang15sbp}.
This literature considers the popularity of pieces of information in isolation. Market-like environments in which \emph{many} messages compete for limited attention have received less consideration. Exceptions have considered the cost of learning about quality~\cite{Adler85}, distortions of quality assessments that result from aggregate knowledge of peer choices~\cite{Salganik854}, and confirmation bias in the spread of misinformation~\cite{DelVicario19012016}. 
The \SimSoM{} model proposed here extends the model of Weng \textit{et al.}~\cite{Lilian2012srep}, who demonstrated that some posts inevitably achieve viral popularity irrespective of quality in the presence of competition among networked agents with limited attention. The model was formalized as a critical branching process and studied analytically, predicting that the popularity of posts follows a power-law distribution with heavy tails~\cite{refgleeson,Gleeson_PRX,Notarmuzi_2018}.  

The current \SimSoM{} model has several limitations. 
First, it neglects many mechanisms of actual online social media that may contribute to message exposure, such as search, personalization, and other details of secret platform algorithms. 
For example, increasingly popular feed ranking algorithms are based less on what is shared by social connections and more on out-of-network recommendations. 
Bad-actor tactics could be affected differently by such algorithmic changes (see \SI). 
While the present model captures several universal criteria of feed algorithms (appeal, social engagement, recency), future work should investigate the vulnerability of different algorithmic affordances. 
Second, real user behaviors are driven by complex cognitive processes. 
Beyond finite attention, such cognitive aspects of information sharing are not explored here. 
Instead, the model assumes that all authentic agents follow a simple probabilistic rule in selecting the content to be shared. 
Large language models have recently been proposed as a way to model agents with more realistic behaviors~\cite{tornberg2023simulating}. 
In addition, \SimSoM{} models information diffusion through simple contagion. While this approach is in line with prior modeling~\cite{bak2022combining}, theoretical~\cite{Juul2021}, and empirical~\cite{hodas2014simple} studies, complex contagion~\cite{centola2018experimental} can also play a role in the spread of certain types of harmful information~\cite{monsted2017evidence}.  
Despite these limitations, the current model's predictions are consistent with empirical findings about the difference in virality between low- and high-quality information on 
Twitter~\cite{Vosoughi1146,Shao18hoaxybots}.
This suggests a reasonable balance between model realism and generality. 

We can think of inauthentic accounts as zealots, a minority of agents committed to a particular view. Opinion dynamics models have shown that a critical minority of active zealots can quickly drive a system to consensus toward their opinion~\cite{galam2007role,PhysRevE.91.022811,PhysRevE.84.011130,PhysRevE.92.042805}. 
In our social media model, we only explore the capacity to suppress information quality rather than to drive consensus to a particular opinion. 

While the structure of online social networks evolves over time, the follower network used in our experiments provides a more realistic setting than synthetic networks to run our simulations. In particular, it captures features of real social networks ---hubs, clustering, and political homophily structure--- that have been observed consistently over the years and across platforms~\cite{conover12partisan,Nikolov2020partisanship,cinelli2021echo} and that are known to play key roles in information diffusion~\cite{Truthy_icwsm2011politics,Lilian2013srep,nematzadeh2014optimal}. Furthermore, our sampling procedure captures a community of accounts that are both active and vulnerable to misinformation, as needed to study manipulation tactics that target such communities. Finally, the assumption that the follower network is static during an information-spreading simulation is consistent with observations that the dynamics of unfollowing are slow~\cite{ashkinaze2024unfollowing}.

Our results suggest that, surprisingly, inauthentic accounts do not need to target hubs (influentials); they can do more damage by connecting to random accounts.
Future research should focus on whether the strategic placement of bad actors within/across polarized online communities can sway users toward a particular outcome, for example by distorting popularity  perceptions~\cite{gerrymandering2019,LermanParadoxDirected2020}. 
Further work is also needed to characterize the effects of strategic targeting when the attacker has limited resources. 

The insights gained from the present findings suggest several countermeasures to increase the resilience of social media to manipulation. The first lesson is that we must make it more difficult for bad actors to infiltrate the network. Platform efforts to detect and take down deceptive accounts must be strengthened, especially as tactics to hack follower networks get more sophisticated~\cite{torres2022trains}.

Recent models show that caps on the depth and/or breadth of diffusion networks can decrease the ratio of distorted messages received by social media users~\cite{Jackson2022_depth_breadth_caps}. While these models assume that distortions occur randomly, \SimSoM{} demonstrates how adversarial actors can exploit vulnerabilities by flooding our news feeds, thereby crowding out quality information.   
A countermeasure would be to 
challenge accounts that post at very high rates 
to prove that they are human. 
Users could also be warned when they follow accounts that post low-quality content and/or when friend accounts are suspended or take suspicious actions, such as changing names/handles~\cite{Pacheco2021Coordinated} or posting and deleting large volumes of content~\cite{Torres2022deletions}. 

Our findings suggest that a winning tactic for malicious accounts is to target all social media users, and not only the most influential, if their goal is to spread harm widely across the network. 
Literacy programs may provide some protection against disinformation. 
This is also suggested by a binary agreement model that assumes each agent may be committed to truth or disinformation~\cite{butts2023mathematical}. 
Social media platforms could lead by educating users about their vulnerability to manipulation and deception. The simplest version of this is through accuracy reminders, which can improve content quality during both posting~\cite{Katsaros_Yang_Fratamico_2022} and resharing~\cite{pennycook2020fighting}. 
Flooding reduces the effectiveness of such accuracy interventions~\cite{Fazio2020pause}. 
Our model could be extended to explore ways to combine friction and accuracy nudges by, e.g., lowering the probability that low-quality messages are reshared by agents after they are exposed to warnings. 

Given that inauthentic accounts can be used to suppress human speech, granting them (or any entity that controls them) unlimited free-speech rights would seem to lead to a logical contradiction~\cite{VanAlstyneCACM2019}. Yet, ironically, efforts by social media platforms to moderate abusive accounts and even research on social bot detection have been assailed by some with charges of censorship~\cite{Mervis2014}. 
These questions 
may have significant repercussions on regulations designed to protect the online public square~\cite{lin2019organization}. 

\section*{Methods} 

\subsection*{Social Media Diffusion Model} 

\SimSoM{}\footnote{Code and data to implement the model and reproduce results are available at \url{github.com/osome-iu/SimSoM}} is a parsimonious agent-based model inspired by the long tradition of representing the spread of ideas as an epidemic process where messages are passed along the edges of a network~\cite{daley}. The model simulates a directed follower network, as in Twitter/X, Mastodon, or Threads. Nodes represent agents (users) and links represent follower relations, which may or may not be reciprocal. The direction of a link goes from the follower to the followed (friend) account, capturing the flow of attention; when a friend's post is reshared by a follower, information spreads in the opposite direction. 
In line with previous work~\cite{hodas2014simple, Juul2021, bak2022combining}, the diffusion process is modeled as simple contagion, where each exposure to a message results in the same resharing probability.
We assume that the structure of the network is static (no unfollowing/blocking of accounts) during the information-spreading process. In contrast to classical epidemiological models, new messages are continuously introduced into the system in an exogenous fashion.

At each time step, an agent $i$ produces a new message with probability $\mu$ or chooses one of the messages in their news feed to be reshared with probability $1-\mu$ (Fig.~\ref{fig:model}). The new or reshared message is then added to the news feeds of $i$'s followers. 
We set $\mu=0.5$, reflecting the empirical average ratio for English-language tweets~\cite{alshaabi2021growing}.
Based on empirical data, agents are assumed to have limited-size inventories: only the most recent $\sigma=15$ messages are retained in each news feed (see \SI). 
The results are robust for different values of $\mu$ and $\sigma$ (see Fig.~\ref{fig:sigmamu} in \SI). 

We assume authentic agents prefer to reshare messages posted by their friends that are appealing, recent, and popular. This is based on empirical evidence that users are more likely to share popular content according to engagement signals~\cite{Fakey2020}. 
The probability that an agent shares a message from their news feed, allowing it to spread, is proportional to the message's appeal, social engagement, and recency. More explicitly, let $M_i$ be the feed of $i$ ($|M_i|=\sigma$). The probability of message $m \in M_i$ being selected is $P(m) = a_m e_m r_m/\sum_{j \in M_i} a_je_jr_j$ where $a_m$ is the appeal of message $m$, $e_m$ is the social engagement, i.e., the number of times it was (re)shared by $i$'s friends, and $r_m$ is the recency. Message recency decays with time as a stretched exponential function $r_{m}(t) = e^{-0.4 t^{0.4}}$, where $t$ is the ``age,'' or the time passed since $m$ was first introduced to the agent's news feed. This decay function is based on empirical online news engagement data~\cite{wuhuberman2007}. 
To model flooding, bad actors can be more active than authentic agents, sharing a message $\theta$ times such that it appears to have higher engagement than authentic content. 

\subsection*{Quality and Appeal}

Both the quality $q$ and the appeal $a$ of new messages are defined in the unit interval. Informed by empirical data, $q$ and $a$ for authentic accounts are assumed to be independent (see Fig.~\ref{fig:corr_appeal_quality} in \SI). For authentic accounts, quality $q$ is drawn from an exponential distribution $P(q) = Ce^{\tau q}$ where high-quality information is more common than low-quality information. The term $C = \frac{\tau}{e^{\tau}-1}$ is a normalizing constant such that $\int_0^1 P(q)dq=1$; the exponent $\tau=10$ is estimated empirically (see Fig.~\ref{fig:quality_func} in \SI). We independently draw appeal from the distribution $P(a) = (1+\alpha)(1-a)^\alpha$, with $\alpha>1$ (Fig.~\ref{fig:appeal_func}(a) in \SI). We set $\alpha = 4$. This choice for the appeal probability density function reproduces a broad distribution of reshares comparable to that observed in real-world information diffusion networks: few messages go viral while the majority do not (Fig.~\ref{fig:appeal_func}(b) in \SI). 

On the other hand, we assume that bad actors strictly generate low-quality messages ($q = 0$). 
The potentially deceptive nature of this content is modeled by the deception parameter $\phi$ ($0 \leq \phi \leq 1$), the probability that a bad-actor message is irresistibly appealing. With probability $\phi$, we set $a=1$, and with probability $1-\phi$ we draw appeal from the same distribution as for authentic accounts, $P(a) = (1+\alpha)(1-a)^\alpha$. 
If $\phi=0$, bad actors and authentic accounts generate messages with appeal drawn from the same distribution. If $\phi>0$, bad-actor messages are more likely to have high appeal; the larger $\phi$, the greater the potential virality of low-quality content by bad actors. 

\subsection*{Follower Network}

We run simulations on a follower network derived from empirical Twitter data.
This network was constructed from a 10\% random sample of public tweets between June 1--30, 2017~\cite{Nikolov2020dataset}. 
The data includes users (excluding likely automated accounts) that shared at least ten links to news sources, at least one of which was to a source labeled as low-quality. Only news sources with known political valence were considered. 
Based on the shared links, most accounts in the dataset have an associated \emph{partisanship score} (from $-1$ for left-leaning to $+1$ for right-leaning) defined as the average political bias of the news sources they share; and a \emph{misinformation score} defined as the fraction of posts linking to low-credibility sources. Both partisanship and misinformation scores, included in the original network dataset, were based on news source labels from third-party fact-checkers~\cite{Nikolov2020partisanship}.

From the original dataset, we select nodes with both partisanship and misinformation attributes. We further reduce the size of the network to speed up our simulations. We apply $k$-core decomposition to select $N = 10{,}006$ nodes forming the $k = 94$ core. Finally, we remove a random sample of edges to decrease the density of this core while preserving the average in/out-degree (number of friends/followers) of the original network ($k=0$ core). This results in $E=1{,}809{,}798$ edges; each node has on average approximately 180 friends/followers (Fig.~\ref{fig:shuffle}, ``hubs+clusters" network). 

\subsection*{Bad-Actor Subnetwork}

Since the empirical network described above does not include likely inauthentic accounts, we use it to model the subnetwork of authentic accounts. 
We then add a subnetwork representing inauthentic accounts that infiltrate the system. The ratio between the sizes of the two subnetworks is described by $\beta$, i.e., for an authentic agent subnetwork of $N$ nodes, the inauthentic subnetwork is composed of $\beta N$ nodes (Fig.~\ref{fig:infiltration}). Since there are many types of inauthentic accounts (trolls, social bots, cyborgs), estimating the percentage of these on social media is a very difficult task. We thus set $\beta=0.05$ following a rough estimation of the prevalence of bots on Twitter.\footnote{\url{theconversation.com/how-many-bots-are-on-twitter-the-question-is-difficult-to-answer-and-misses-the-point-183425}} Note that from the perspective of information quality in the model, increasing the prevalence of bad actors is equivalent to increasing their infiltration. Therefore we focus on the effects of varying $\gamma$ rather than $\beta$.

The inauthentic account subnetwork is created using a directed variant of the random-walk growth model~\cite{vazquez03Growing}. This captures the presence of hubs and clustering (directed triads). Specifically, the network is initialized with four fully connected nodes. We then add one new node at a time, assuming that each has fixed out-degree  $k_{out}=3$. Once a new node $i$ comes into the network, it links to (follows) a randomly selected target node (friend) $j$. Each of the remaining $k_{out} - 1 = 2$ friends are selected as follows: with probability expressed by a parameter $p$, $i$ follows a random friend of $j$'s; with probability $1-p$, $i$ follows another randomly selected node. Following friends of a friend has the effect of generating closed, directed triads and approximates a preferential attachment process, giving rise to hub nodes with high in-degree. The parameter $p$ thus models both hubs and clustering.  We use $p=0.5$. 

In addition, the bad-actor subnetwork is designed to manipulate information flow and collective attention by spreading certain messages. Therefore we assume that bad actors get random authentic accounts to follow them: we add a directed link from each authentic node to each bad-actor node with probability $\gamma$. (In the next subsection we present alternatives to this random targeting tactic.)
The parameter $\gamma$ models the degree of infiltration of the network by bad actors (Fig.~\ref{fig:infiltration}). When $\gamma=0$, there is no infiltration and bad actors are isolated, therefore harmless; the opposite extreme $\gamma=1$ indicates complete infiltration such that bad actors dominate the network. 
Because we are not concerned with the quality of messages on bad-actor news feeds, they do not follow or reshare content from authentic agents for the analyses in this paper.  

\subsection*{Bad-Actor Targeting Tactics}

In the scheme described above, the authentic agents that follow bad actors do so randomly. We also study scenarios in which bad actors target certain accounts as potential followers (Fig.~\ref{fig:targeting}). In all of these scenarios, each bad actor is still followed by $\gamma N$ authentic accounts on average, but these targets are selected according to some criterion, such as whether they are politically active or have many followers. 

Each targeting tactic is modeled by making the probability that an authentic agent $i$ follows bad actors proportional to some feature $f(i)$.
In the hubs tactic, we set $f(i)=k_{in}(i)$, the number of followers of $i$.
Misinformation and partisanship attributes of authentic accounts in the empirical network allow us to model other targeting scenarios. In the misinformation tactic, $f(i)$ is set to $i$'s misinformation score, i.e., their propensity to share misinformation. The partisanship score is used in the liberal and conservative tactics, while its absolute value is used in the partisanship tactic. 

\subsection*{Overall Quality} 

The effects of manipulation on authentic agents are quantified by the quality of all content in circulation. The overall quality of the information system at time $t$ is measured by the \emph{average quality} across all the messages visible through the feeds of the authentic agents: 
\[
Q_t = \frac{1}{\sigma N} \sum_{i=1}^N\sum_{m \in M_i} q_{i,m,t},
\]
where $q_{i,m,t}$ is the quality of the message in the $m$-th position in authentic user $i$'s feed at time $t$.

The system's quality at each time step $t$ is calculated with an exponential moving average $\bar{Q}_t = \rho \bar{Q}_{t-1} + (1-\rho) Q_{t}$. 
As the simulation takes place, some of the messages become obsolete quickly, while others live longer and infect a larger fraction of the network. The simulation ends once the system reaches a \emph{steady state} in which the difference between two consecutive values of the quality moving average is smaller than a threshold, i.e., $|\bar{Q}_t - \bar{Q}_{t-1}| / \bar{Q}_{t-1} < \epsilon$. See \SI{} for further details about model convergence to a steady state. 
The average quality reported for all analyses is calculated at the end of the simulation.

\subsection*{Cascade Construction} 

A \emph{reshare cascade} is a tree that begins when an agent $i$ (root) posts a new message $m$. 
After that, a new link is created when a follower $j$ of $i$ reshares $m$. We say that $i$ is the parent of $j$ in the tree. 
Similarly, an agent's exposure to a message is defined as having that message on their feed while being activated --- we assume that an agent who is about to share or reshare a post has seen the content on their feed. 
A link in the \emph{exposure cascade} for message $m$ is created between an activated agent with $m$ in their feed and their friend who had shared $m$. 
When constructing both reshare and exposure cascades, $m$ may have been reshared by more than one of an agent's friends. When this occurs, one of them is selected at random to be the parent in the cascade tree. 
This choice does not affect a cascade tree's size, but might affect its structure. Other alternatives, such as selecting as parent the friend who most recently shared $m$, can be explored if one desires to analyze the structure of the reshare or exposure cascades. 
By definition, reshare cascades can be as small as one and exposure cascades can be as small as zero. 

To capture the complete diffusion cascades, the size distributions plotted in Fig.~\ref{fig:cascade-distributions} include only cascades of ``extinct" messages, those that have become obsolete and are no longer present in any news feed by the end of simulations.

\subsection*{Simulation Framework and Parameters} 

For each set of parameters, we run 10 simulations starting from random conditions; the reported average quality is the average across these runs. 

We explore values of the bad-actor parameters to cover broad ranges of manipulation tactics. The infiltration parameter $\gamma$ is a probability, and we consider values in the unit interval on a logarithmic scale to focus on more realistic low-probability values. The deception parameter $\phi$ also represents a probability and we examine the full unit interval. Finally, the flooding parameter $\theta$ is a multiplicative factor on the volume of posts, and we explore the range 1--64 on a logarithmic scale to capture the high volumes observed in empirical analyses~\cite{Torres2022deletions}. 
In scenarios with bad actors in the system, the default parameters are $\gamma=0.01$, $\phi=0$, and $\theta=1$. Except for the simulations exploring inauthentic targeting tactics, authentic followers of the bad actors are selected at random.

The parameters $\rho =0.8$ and $\epsilon = 0.0001$ were tested to ensure that the system's quality stabilizes at the steady state (see \SI). 

We shuffle the follower network in various ways to derive the scenarios reported in Fig.~\ref{fig:shuffle}. The ``hubs+cluters" network is the original one. In the ``hubs" network, we shuffle the original network while preserving the degree distribution and destroying any community clustering. In the ``clusters" network, the outgoing endpoint of each edge in the original network is rewired to another node within the same cluster; this preserves the cluster structure and destroys the hub structure. In the ``random" shuffle, all edges are rewired at random with uniform probability, destroying all hub and clustering structure. 

\section*{Data Availability Statement} 

Code and data to implement the model and reproduce results are available at \url{github.com/osome-iu/SimSoM}. 

\section*{Acknowledgments}

We are grateful to Marshall Van Alstyne, David Axelrod, Rachith Aiyappa and Erfan Samieyan for useful discussion and suggestions; to Dimitar Nikolov, Chengcheng Shao, Giovanni Ciampaglia, and Pik-Mai Hui for the data collection used to construct the empirical network; to Kai-Cheng Yang and Christopher Torres-Lugo for the COVID-19 data collection; and to Alireza Sahami Shirazi for data about scrolling session depth on a social media mobile app. 
This work was carried out in part while XL visited the Observatory on Social Media at Indiana University, with support by the China Scholarship Council. 
BT, AF and FM were supported in part by The Knight Foundation. 
BT was funded in part by the Ostrom Workshop. 
AF and FM were funded in part by DARPA (grants W911NF-17-C-0094 and HR001121C0169). 
FM and BT were supported in part by the Swiss National Science Foundation (Sinergia grant CRSII5\_209250).
FM was supported in part by Democracy Fund and Craig Newmark Philanthropies. 
Any opinions, findings, and conclusions or recommendations expressed in this material are those of the authors and do not necessarily reflect the views of the funding agencies.

\bibliographystyle{unsrt}
\bibliography{simsom.bib}

\newpage

\setcounter{page}{1}
\renewcommand\thepage{S\arabic{page}}

\appendix

\renewcommand\thesubsection{S\arabic{subsection}}

\counterwithin{figure}{section}
\counterwithin{table}{section}

\counterwithout{figure}{section}
\counterwithout{table}{section}

\renewcommand{\thefigure}{S\arabic{figure}}
\renewcommand{\thetable}{S\arabic{table}}

\setcounter{figure}{0}
\setcounter{table}{0}



\section*{\SI}

\subsection{News feed size and information load}

We use a news feed size $\sigma=15$. This value is approximated from the average depth of approximately 107 mobile scrolling sessions measured on Tumblr during two weeks in 2016. The feed interface of this app was similar to those of other social media platforms. We considered a session to have ended when there was no interaction for 30 minutes or longer. The depth of a session was recorded as the number of times that a user scrolled at least 500 pixels through the feed and then stopped for at least one second. 

\begin{figure}[b!]
\centerline{\includegraphics[width=\normalwidth]{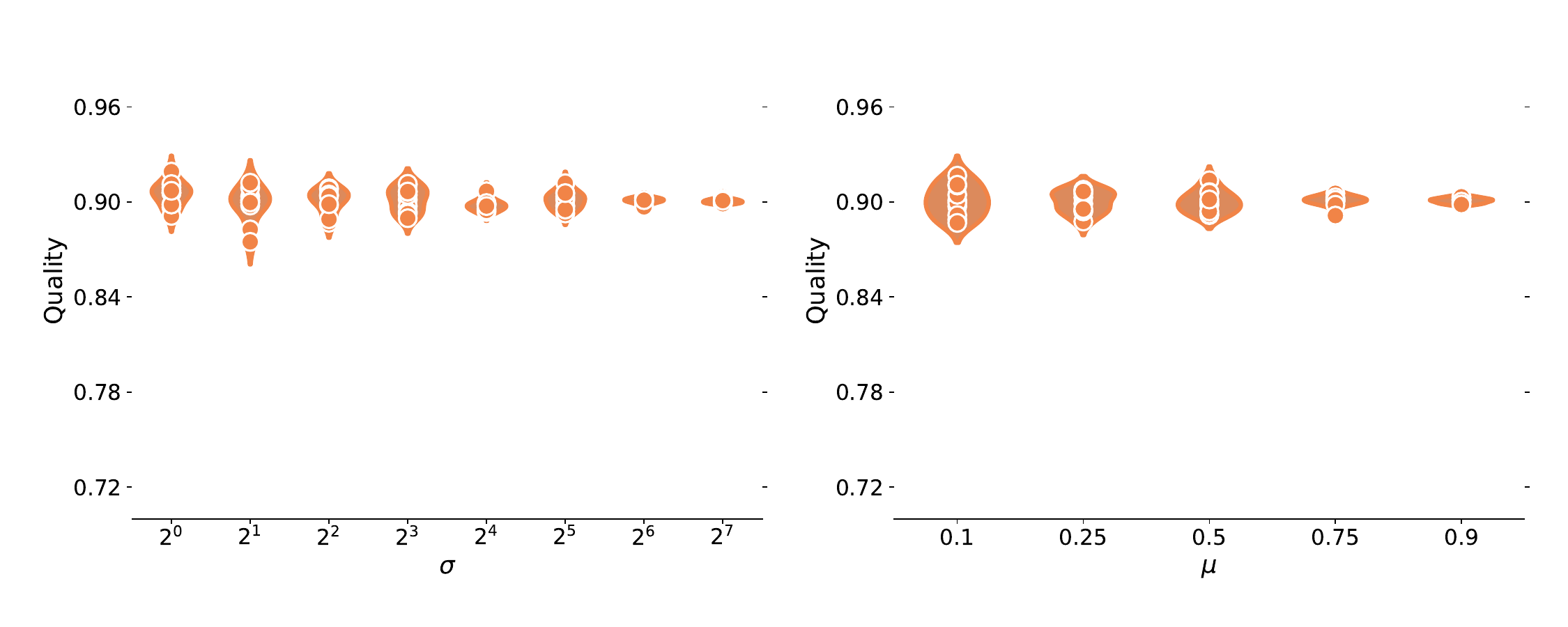}}
\caption{Effects of cognitive features of authentic agents on information quality. Left: News feed size $\sigma$. Right: Information load $\mu$. The simulations are run on the original network without bad actors. The results for $\mu=0.5$ and $\sigma=15$ are used as a baseline to calculate relative quality in the main text.}
\label{fig:sigmamu}
\end{figure}

Results are robust for different values of news feed size $\sigma$ as well as posting activity $\mu$  (Fig.~\ref{fig:sigmamu}).

\subsection{Relationship between quality and engagement}

To estimate the relationship between quality and engagement, we start from a dataset of tweets about COVID-19~\cite{Yang2020covid}. 
The dataset consists of tweets containing the hashtags \#coronavirus or \#covid19 collected in 2020, selected from a 10\% random sample of public tweets~\cite{osome}.
From this set of tweets, we analyze only posts that shared links to news sources from March 9--29, 2020.  
Linked sources were extracted from URLs in the tweet metadata. URLs shortened with 70 popular URL shortening services were also expanded to reveal the actual sources. Posts with links to Twitter and other social media platforms were excluded. 
The remaining links were matched against sources with ratings obtained from NewsGuard\footnote{\url{newsguardtech.com}} in April 2021. 
The final dataset contains 110,224 original Twitter posts that share at least one link with a NewsGuard rating. For each post, we have both the rating of the shared source, used as a proxy for post quality, and the number of retweets. The latter is extracted from the metadata of the latest retweet of each post.

\begin{figure}[t!]
\centerline{\includegraphics[width=\morereducedwidth]{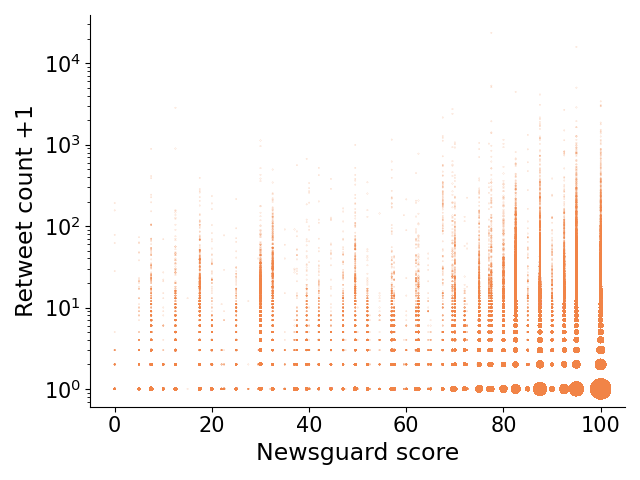}}
\caption{Scatter plot of retweet counts and Newsguard scores of original tweets with links to news articles. The size of each point is proportional to the number of tweets having a given retweet count and Newsguard score. We add one to the retweet count so that the tweets with zero retweets are visible on the log scale. The slightly negative correlation is driven by the majority of tweets linking to high-quality sources and having no retweets.}
\label{fig:corr_appeal_quality}
\end{figure}

Fig.~\ref{fig:corr_appeal_quality} shows that the engagement of a post, as measured by its retweet count, is weakly correlated with its quality (Spearman coefficient $-0.13$). Based on this observation, we assume that appeal is independent of quality in the model.

\subsection{Empirical estimation of quality distribution}

\begin{figure}[b!]
\centerline{\includegraphics[width=\morereducedwidth]{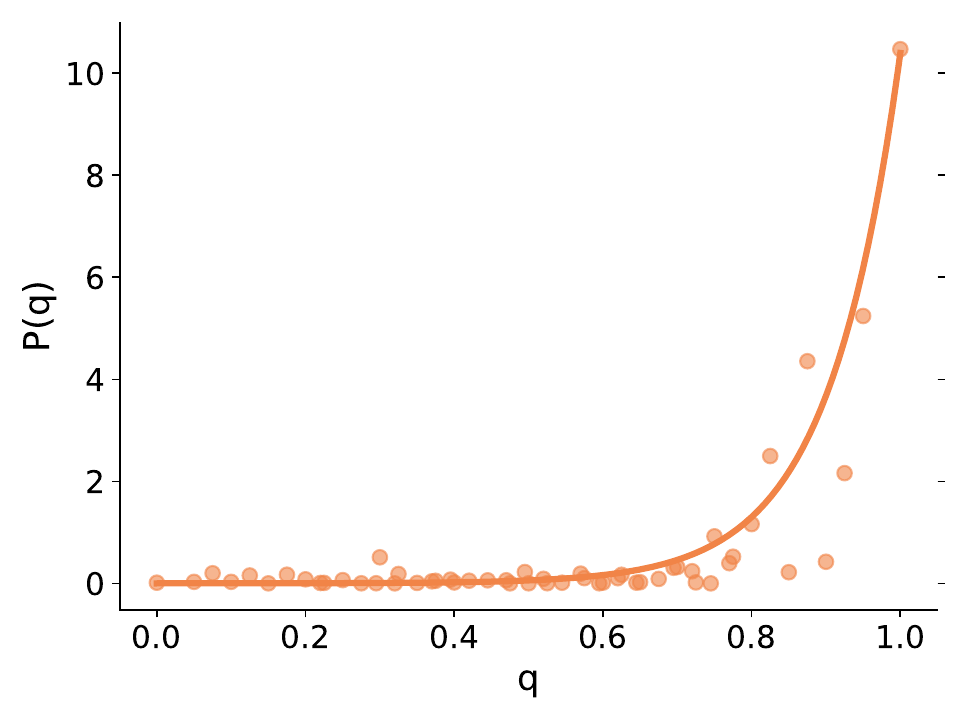}}
\caption{Distribution of quality for authentic agent messages (line) fitted to empirical Twitter data (points).}
\label{fig:quality_func}
\end{figure}

The same empirical data described above is used to estimate the distribution of authentic message quality. The quality of a post is the average Newsguard score of the news sources shared in the post, normalized to be in the unit interval. As shown in Fig.~\ref{fig:quality_func}, we fit the data to the exponential probability density function $P(q) = \frac{\tau}{e^{\tau}-1}e^{\tau q}$, with $\tau=10$. The normalization constant is obtained by setting $\int_0^1 P(q)dq=1$.

\subsection{Empirical estimation of appeal distribution}

\begin{figure}
\centerline{\includegraphics[width=\normalwidth]{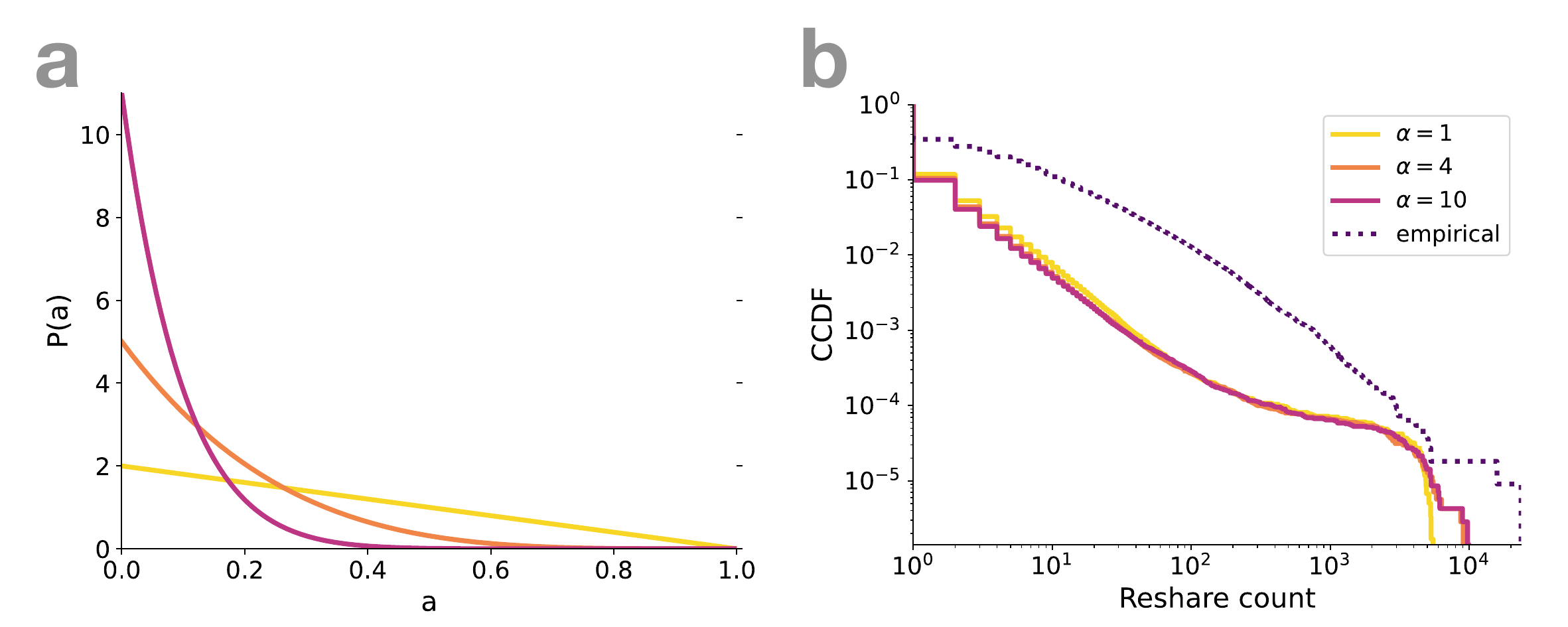}}
\caption{Appeal distribution. (a) Distribution of appeal for authentic agent messages, for different values of the parameter $\alpha$.  
(b) Reshare count distribution of messages in the baseline simulation of the model with only authentic accounts, and of empirical posts. The message reshare counts in the model are measured at the steady state.}
\label{fig:appeal_func}
\end{figure}

The appeal distribution of authentic content is modeled by the probability density function $P(a) = (1+\alpha)(1-a)^\alpha$. 
The normalization constant is obtained by setting $\int_0^1 P(a)da=1$.
This captures the intuition that most messages have low appeal. 
As illustrated in Fig.~\ref{fig:appeal_func}(a), high appeal values are much more rare for $\alpha=10$ than $\alpha=1$. 
While there is no empirical proxy data for appeal, a reasonable choice should give rise to a distribution of engagement consistent with empirical reshare data. 
Fig.~\ref{fig:appeal_func}(b) shows that values $\alpha \geq 1$ result in distributions of engagement in the model that roughly capture the broad distribution of reshares in the same empirical data described above. 

\subsection{Convergence to steady state}

\begin{figure}
\centerline{\includegraphics[width=\normalwidth]{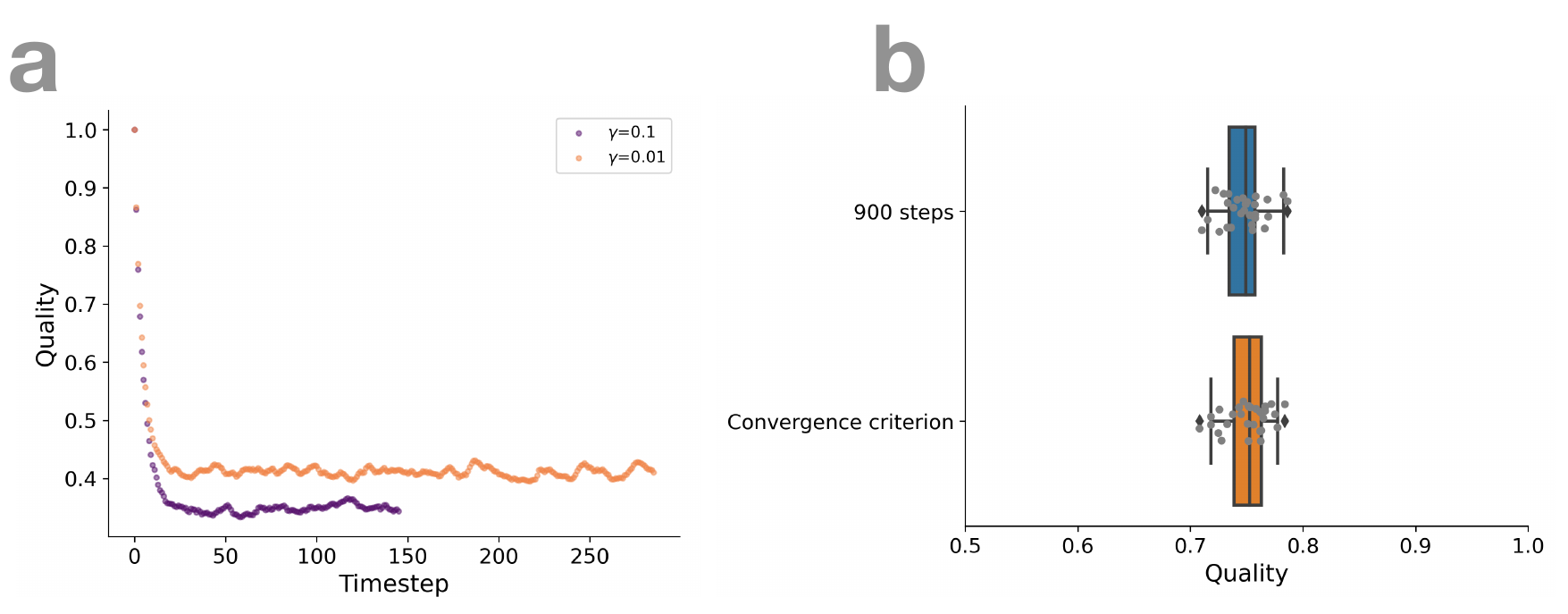}}
\caption{(a) Overall quality across time steps for two simulation runs with different infiltration values. 
(b) Average quality at the end of prolonged simulations vs.~when our convergence criterion is met. Plots represent the median (center bar), 50\% (box) and 95\% (whiskers) confidence intervals, and outliers (diamonds). The actual values from the simulations are represented as gray dots.}
\label{fig:convergence}
\end{figure}

Fig.~\ref{fig:convergence}(a) illustrates how the system reaches a steady state: after a transient period, the overall quality $\bar{Q}_t$ stabilizes around a consistent average value. 
To ensure that the overall quality of information $\bar{Q}$ is stable when the convergence criterion is met, we tested many values of the parameters $\rho$ and $\epsilon$. The time it takes to converge varies with different simulation parameters, like $\gamma$, $\phi$, and the population size. Using $\rho=0.8$ and $\epsilon=0.0001$, we observe that when the criterion is met, $\bar{Q}$ is stable for various values of the other parameters. Thus, we use these values in all the experiments reported in the main text.

Additionally, we ran 20 simulations for an extended duration of 900 time steps, well beyond the typical convergence point. Fig.~\ref{fig:convergence}(b) compares the average quality $Q$ when the convergence criterion is met with that at the end of the extended simulations. The average quality values are similarly distributed in the two conditions, indicating stability. 
Note that the fluctuations of the overall quality $\bar{Q}$ are much smaller.

\subsection{Alternative feed algorithm}

The model analyzed in the manuscript assumes that users are exposed to content (inventory) only form the accounts they follow. While this was typical of many social media platforms until recent years, in the wake of TikTok's success several platforms now employ recommendation-based algorithms that mix in inventory from the entire network. While a full investigation of alternative algorithms is outside the scope of the present manuscript, we carried out a limited analysis to explore the effects of this kind of variations on our main results. 

\begin{figure}[t]
\centerline{\includegraphics[width=\normalwidth]{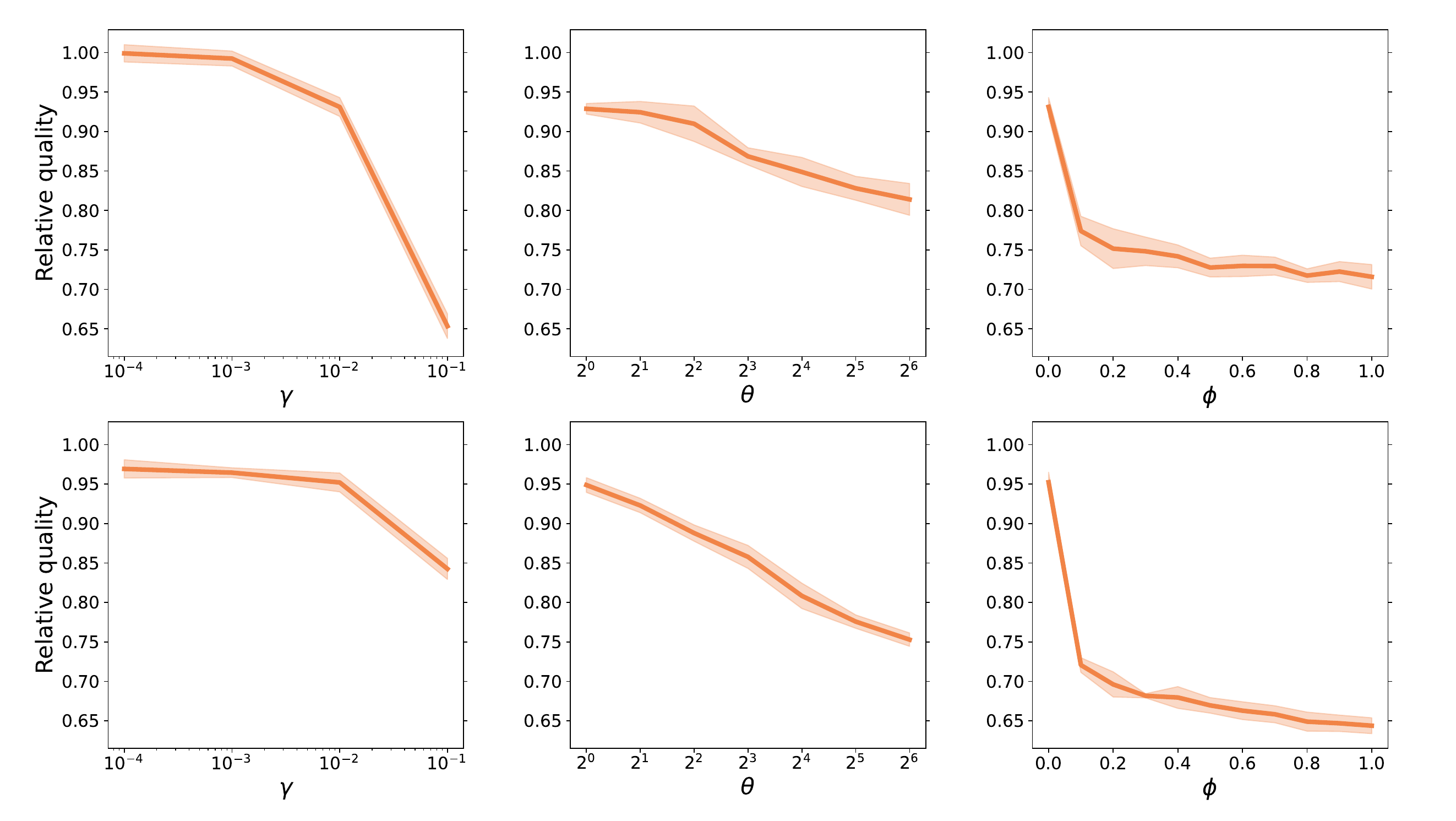}}
\caption{Comparison between the effects of individual bad-actor tactics on message quality, relative to the scenario without bad actors, with different feed algorithms. 
The top row shows results with the original algorithm (in-network inventory); the bottom row shows results with the alternative algorithm (50\% in-network, 50\% out-network). 
From left to right we report on the effect of varying infiltration $\gamma$ ($\theta=1$, $\phi$=0);
flooding $\theta$ ($\gamma=0.01$, $\phi=0$); and
deception $\phi$ ($\gamma=0.01$,  $\theta=1$). 
Shading represents 95\% confidence intervals across runs.}
\label{fig:supp:robustness}
\end{figure}

We consider a scenario in which 50\% of each agent's feed inventory comes from in-network posts (by friend accounts) and 50\% from out-network posts (by the rest of the accounts). We repeat the analyses in Fig.~\ref{fig:bottactics}(a,b,c), exploring the effects of bad-actor tactics based on infiltration, flooding and deception. We employ a smaller network obtained by down-sampling the follower network used in the main text ($N=1{,}000$ nodes, $E=30{,}030$ edges, 30 friends/followers per node on average). The results, shown in Fig.~\ref{fig:supp:robustness}, are qualitatively similar to those obtained using the original algorithm based on only in-network inventory. However, infiltration has less impact while flooding and deception are more harmful in the mixed-inventory scenario. 
These changes can be understood by the observations that bad actors also receive and spread content from authentic agents, making their infiltration less harmful, while high-volume, high-appeal low-quality content can now spread farther.  
This analysis suggests that the harm caused by different bad-actor tactics can be mitigated or amplified by algorithmic affordances. 

\end{document}